\documentclass{article}

\usepackage{graphicx}
\usepackage{dcolumn}
\usepackage{bm}
\usepackage{color}
\usepackage{lscape}
\usepackage{amscd,amsmath,amssymb}
\newcommand{\be}{\begin{equation}}
\newcommand{\bel}{\begin{equation}\label}
\newcommand{\ee}{\end{equation}}
\newtheorem{proposition}{\bf Proposition}[section]

\newlength{\intwidth}

\begin{document}

\title{Self-similarity in turbulence and its applications}

\author{
  Koji Ohkitani\\
  Research Institute for Mathematical Sciences,\\
  Kyoto University,  Kyoto 606-8502 Japan.
}




\maketitle
\begin{abstract}

  
  \textcolor{black}{First, we discuss the non-Gaussian type of self-similar solutions to the Navier-Stokes equations.}
  We revisit a class of self-similar solutions which was  studied in Canonne-Planchon (1996).
  In order to shed some light on it, we study self-similar solutions to the 1D Burgers equation in detail,
  completing the most general form of similarity profiles that it can possibly possess.
  In particular, on top of the well-known source-type solution we identify a kink-type solution.
  It is represented by one of the  confluent hypergeometric functions, viz. Kummer's function $M.$

  For the 2D Navier-Stokes equations,  on top of the celebrated  Burgers vortex
  we derive yet another solution to the associated Fokker-Planck equation.
  This can be regarded as  a 'conjugate' to the  Burgers vortex, just like the kink-type solution above.
\textcolor{black}{Some asymptotic properties of this kind of solution have been worked out. 
  Implications for the 3D Navier-Stokes equations are suggested.}

Second, we address an application of self-similar solutions to explore more general kind of
  solutions. In particular, based on the source-type self-similar solution to the 3D
  Navier-Stokes equations, we consider what we could tell about more general solutions. 
  
\end{abstract}


\begin{flushleft}
Burgers equation, Navier-Stokes equations, self-similarity
\end{flushleft}

\section{Introduction}
Self-similarity is an useful concept in handling partial differential equations
particularly arising from fluid mechanics. Our motivation for this study is as follows.

An initial-boundary-value problem for the 3D Navier-Stokes equations is studied in \cite{CP1996}
postulating that the velocity field has a self-similar form, from the very initial data to the final
state. To this end they introduced  Besov spaces to accommodate a singular initial
velocity field which is as rough as  $u(x) \propto 1/x$. The knock-on effect is that we would have
similarity profiles, viz. steady solutions to the scaled
Navier-Stokes equations, which are {\it not} well-localised in space.
In order to shed some light on its implications, we will take up a simpler system, the 1D Burgers
equation to consider solutions in such an enlarged function space.
In particular, we will be concerned with the fate
of solutions which are not in $L^1$marginally, but in some Besov space nearby.

\textcolor{black}{There are two schematic themes in this paper. One is the study of non-Gaussian type solutions
  (loosely called 'kink-type' solutions) of scaled version of fluid dynamical equations. The other one is 
  proposal of a new approach of constructing general solutions on the basis of particular source-type
  self-similar solutions.}

\textcolor{black}{
  For the first theme,  non-Gaussian solutions are given explicitly with some of their properties discussed,
  whereas clarification of their possible significance as a dynamical system is left for further study.
  For the second theme a protocol for (re)building more general solutions out of self-similar solutions
  is exemplified using the Burgers and 2D Navier-Stokes equations. This suggests that applications to
  the 3D Navier-Stokes equations  deserve further investigation.}

This paper is organised as follows. We study self-similar profiles for the Burgers equation in Section 2,
emphasising the properties of the newly identified kink-type solutions.
We consider self-similar solutions  to the Navier-Stokes equations, especially in two dimensions in Section 3.
We address possible applications of those similar solutions, in particular about obtaining information regarding
more general class of solutions in Section 4. Section 5 is devoted to  summary and outlook.

\section{Self-similar solutions of the Burgers equation}

When it is demanded that the initial data themselves are self-similar, inevitably the initial velocity
field goes singular like
$u(x) \propto 1/x$.
This has difficulties both at the origin and at infinity: (1) it is singular and non-integrable
at the origin and (2)  its decay at far distances is too slow to be integrable.
For those reasons, Besov spaces\footnote{They are associated with a norm defined by finite-differences;
  $\|\bm{u}\|_{B^s_{pq}} \equiv \left\{\sum_{j=1}^\infty
\left( 2^{sj} \|\Delta_j (\bm{u}) \|_{L^p}\right)^q \right\}^{1/q},$
where $1 \leq p,q \leq \infty, s \in \mathbb{R}$ and $\Delta_j (\bm{u})$
represents the band-filtered velocity at frequency $2^j$.}  
were introduced to construct solutions
to the initial-boundary problem in \cite{CP1996}.

Assume the initial data is in a Besov space $B^0_{3,\infty}(\mathbb{R}^3)$
but not in $L^3(\mathbb{R}^3)$.
The construction is concerned with a class  of solutions for $t>0,$
which are in $B^0_{3,\infty}(\mathbb{R}^3)$ but not in $L^3(\mathbb{R}^3)$.
The 'source-type' self-similar solutions are irrelevant here, because they are well-localised with
a finite $L^3$-norm.
The obvious question is: which functions can the \textcolor{black}{scaled} solutions possibly tend to, if they approach steady solutions at all ?

To shed some light on solutions constructed in Besov spaces 
we consider a simpler  problem of the 1D Burgers equation.
For this purpose, instead of $B^0_{3,\infty}(\mathbb{R}^3)$ we consider
velocity fields in $B^0_{1,\infty}(\mathbb{R}^1)$  to study an analogous problem
in one spatial dimension.
It is readily checked that the theory developed in \cite{CP1996}
holds valid  \textit{mutatis mutandis} for the 1D Burgers equation.
\textcolor{black}{For more recent references on self-similar solutions, including studies
  with use of BMO-type spaces,   see for example \cite{Rieusset2002, GPS2007, BT2018}.}
  
\subsection{Cole-Hopf transform as Riccati substitution}
We consider the Burgers equation in $\mathbb{R}^1$
\bel{Burgers.eq}
\frac{\partial u}{\partial t} + u \frac{\partial u}{\partial x}
=\nu \frac{\partial^2 u}{\partial x^2},
\ee
for an initial data $u(x,0)=u_0(x)$.
It is well-known that (\ref{Burgers.eq}) is linearisable by the so-called Cole-Hopf transform.
We recall the basic results obtained in \cite{Hopf1950}.
\begin{itemize}
\item If $\int_0^x u_0(y)dy =o(x^2)$ for large $|x|,$ unique existence of solutions is guaranteed
for all $t>0$.

\item If $\int_0^x u_0(y)dy =O(x^2)$ for large $|x|,$ the existence of solutions is guaranteed only
  for finite time
$0 \leq t < T$. 
\end{itemize}
An instructive example for the second result is given by $u(x,t)=\frac{x}{t-T}$ \cite{Hopf1950}.

To discuss forward self-similar solutions we introduce dynamic scaling transformations
$$\xi=\frac{x}{\lambda(t)},\;\tau=\textcolor{black}{\frac{1}{2 a}\log t},$$
$$u(x,t)=\frac{1}{\lambda(t)}U(\xi,\tau),$$
where  \textcolor{black}{$\lambda(t)=\sqrt{2at}$} denotes a scaling length and
$a (>0)$ a zoom-in parameter.
Applying them to (\ref{Burgers.eq}) we find the scaled form of the  Burgers equation
\bel{scaledBurgers}
\frac{\partial U}{\partial \tau} +U \frac{\partial U}{\partial \xi}
=a  \frac{\partial}{\partial \xi}\left( \xi U \right)
+\nu \frac{\partial^2 U}{\partial \xi^2}.
\ee
If a steady solution is established \textcolor{black}{in the self-similar variables}
as $\tau \to \infty,$ it satisfies
$$ \frac{d^2 U}{d \xi^2} + \frac{a}{\nu}\frac{d}{d \xi}\left( \xi U \right)
= \frac{d}{d \xi} \frac{U^2}{2\nu}.$$
After an integration we find  the following form
\bel{Riccati}
\frac{d U}{d \xi}+\frac{a}{\nu} \xi U =\frac{U^2}{2\nu}+2C,
\ee
where the prefactor of $2$  is inserted in front of a constant $C$
for subsequent convenience. (In this paper $c, C, C_1,\ldots$  etc. denote constants
which may be different from line to line.)

\begin{figure}[ht]
\begin{minipage}{0.55\linewidth}
  \includegraphics[scale=0.4,angle=0]{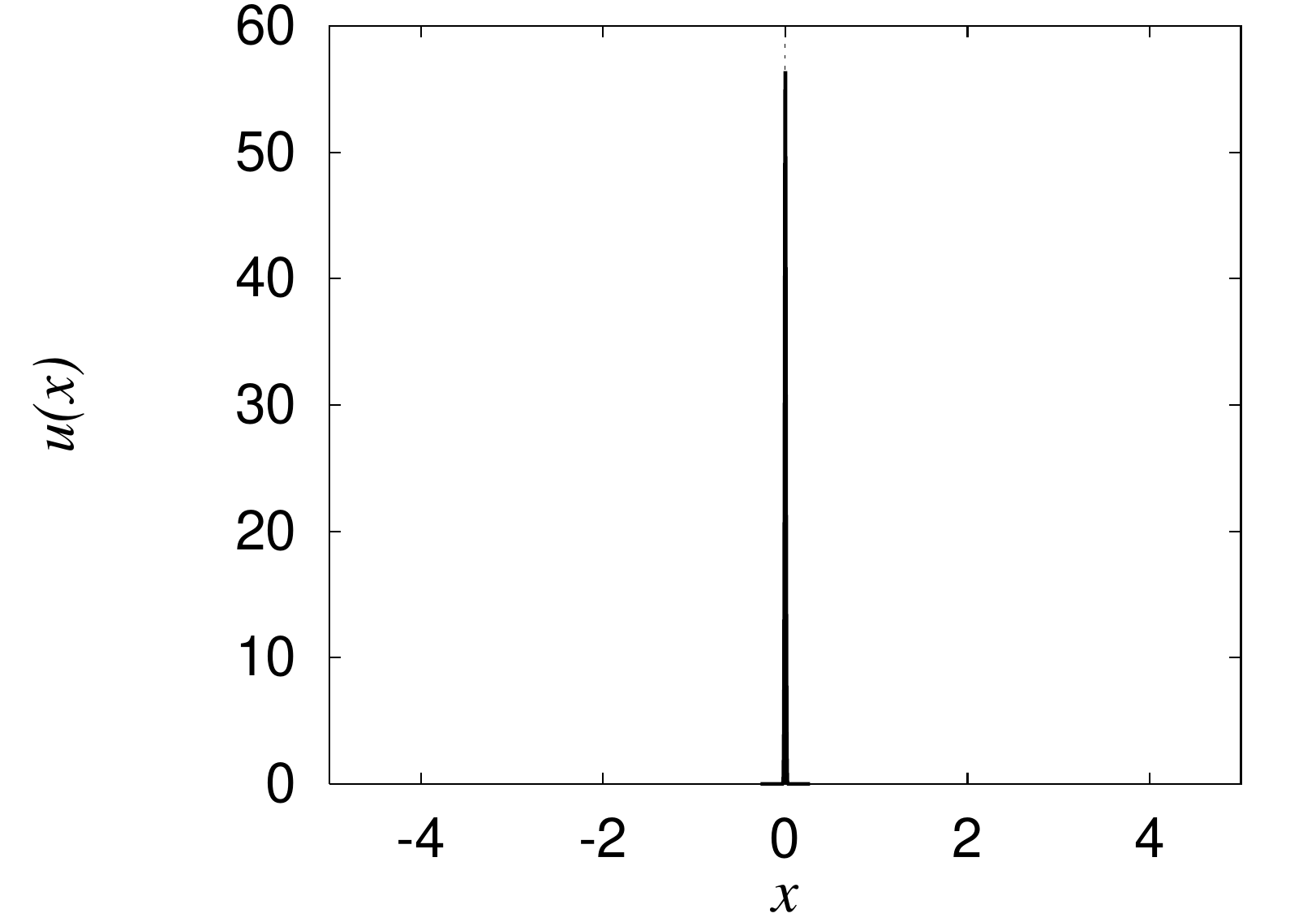}
  \caption{The initial data, Dirac delta function
    \textcolor{black}{$u(x)=\delta(x)$ at $t=0$.} Depicted as $\delta_\epsilon(x)\equiv \exp(-x^2/\epsilon)/\sqrt{\pi \epsilon},$
    with $\epsilon=1\times10^{-4}.$     (The figures in this paper are meant to be schematic.)}
\label{IC_delta}
\end{minipage}
\begin{minipage}{0.5\linewidth}
  \includegraphics[scale=0.25,angle=0]{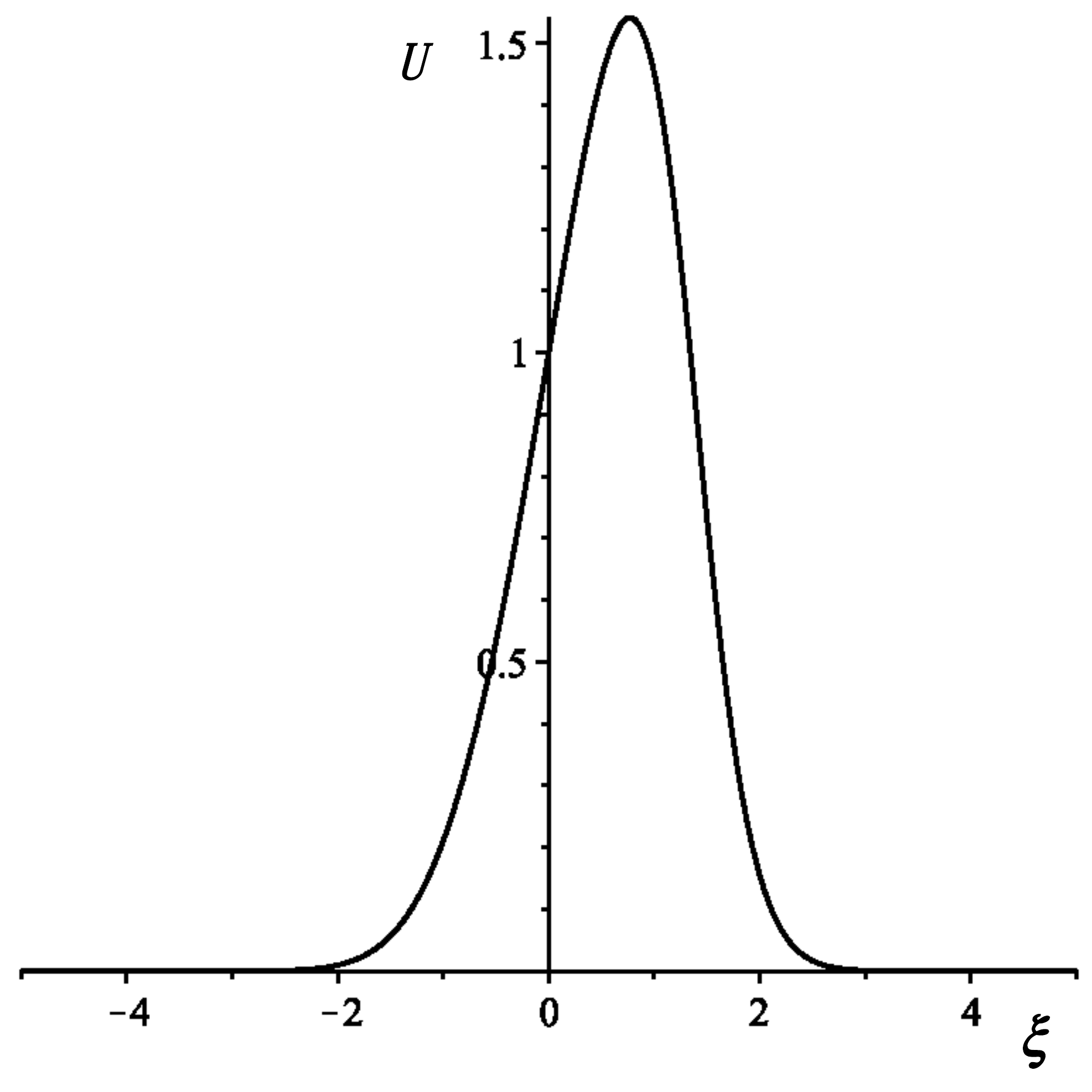}
  \caption{The source-type solution as $\tau \to \infty$;
  $U(\xi)=\dfrac{e^{-\xi^2}}{1-\int_0^\xi e^{-\eta^2}d\eta.}$}
  \label{Burg_source}
\end{minipage}
\end{figure}

We now distinguish  two cases.
\begin{enumerate}
\item    If $C=0,$
 the problem is well-understood. It  can be solved by either introducing an integrating factor
$\frac{d U}{d \xi}+\frac{a}{\nu} \xi U= e^{-\frac{a \xi^2}{2\nu}}
\frac{d}{d\xi}\left(U e^{\frac{a \xi^2}{2\nu}} \right),$
or linearising with a transformation $V=1/U$ regarding it as a Bernoulli equation.
Either way we find the so-called source-type solution,  e.g. \cite{EZ1991, BKW1999}
\bel{Burgers_source1}
U(\xi)=\frac{C' \displaystyle{\exp \left( -\frac{a \xi^2}{2 \nu} \right)}}
{1 -\displaystyle{\frac{C'}{2\nu}\int_0^{\xi}}
  \exp \left( -\frac{a \eta^2}{2 \nu}\right) d\eta},
\ee
for a constant $C'$. The solution is well-localised spatially, see Figs.\ref{IC_delta}, \ref{Burg_source}.
The name has come from the fact that
$\lim_{t \to 0}\frac{1}{\sqrt{2at}} U(\xi)=M\delta(x),$
where $\delta(\cdot)$ denotes the Dirac mass and $M\equiv\int_{-\infty}^{\infty} u_0(x) dx.$
We also recall \cite{EZ1991, BKW1999} that for all $u_0 \in L^1(\mathbb{R}^1)$  and $1 \leq p \leq \infty,$
$$t^{\frac{1}{2}\left(1-\frac{1}{p}\right)}                                   
\left\| u(x,t)- \frac{1}{\sqrt{2at}} U(\xi)\right\|_{L^p}                     
\to 0\;\;\mbox{as}\;\; t \to \infty,$$
showing some degree of the universality of the profile.
\item If $C \ne 0,$ (\ref{Riccati}) is a Riccati equation and
  the method above does \textit{not} work. In this case, we ought to introduce a Riccati substitution
    $U=-2\nu \dfrac{\Psi'}{\Psi},$ 
  to reduce it to a linear equation (see {\bf Appendix  A})
\bel{Riccati2}
\Psi'' +\frac{a}{\nu}\xi \Psi' +\frac{C}{\nu}\Psi=0.
\ee
It is to be observed that the Cole-Hopf transform arises 
  as a natural course of solution to the Riccati equation \cite{DJ1989}.
If the initial data  $u_0 \notin L^1(\mathbb{R}^1),$ but  $u_0 \in B^0_{1,\infty}(\mathbb{R}^1),$
the other kind of solutions, i.e. the kink-type ones will come into play.
  Below we  will discuss those solutions to (\ref{Riccati2}) in detail.
  \end{enumerate}

\subsection{Kink-type solution}
As noted already, the theory developed in \cite{CP1996} for  the 3D Navier-Stokes equations
works for initial velocity $\notin L^{3}(\mathbb{R}^{\textcolor{black}{3}})$, but
$\in B^0_{\textcolor{black}{3},\infty}(\mathbb{R}^3).$ 
We will consider an analogous 1D problem  as an illustration.

In fact, the equation (\ref{Riccati2}) can be solved using confluent hypergeometric functions, viz. the Kummer's
functions. Consider
\bel{conf.hyp.geom}
\Psi'' +A\xi \Psi' +B\Psi=0,
\ee
where $A$ and $B$ are constants (to be set $A=\frac{a}{\nu}$ and $B=\frac{C}{\nu}$).
Actually we have the following
\begin{proposition}
The solution to (\ref{conf.hyp.geom}) can be written
$$\Psi=\xi e^{-\frac{A}{2}\xi^2} w\left(1-\frac{B}{2A},\frac{3}{2},\frac{A}{2}\xi^2\right),$$ 
where the function $w\left(1-\frac{B}{2A},\frac{3}{2},z\right)$ satisfies the confluent
hypergeometric equation
\bel{conf.hyp.geom2}
z\frac{d^2 w}{dz^2}+\left(\frac{3}{2}-z\right)\frac{dw}{dz}+\left(\frac{B}{2A}-1\right) w=0.
\ee
Hence the general solution is given by
$$\Psi=C_1 \xi\exp\left(-\frac{a}{2\nu}\xi^2\right) M_{\rm K}\left(1-\frac{C}{2a},\frac{3}{2},\frac{a}{2\nu}\xi^2\right)
+C_2 \xi\exp\left(-\frac{a}{2\nu}\xi^2\right) U_{\rm K}\left(1-\frac{C}{2a},\frac{3}{2},\frac{a}{2\nu}\xi^2\right),$$
where  $M_{\rm K}(\alpha,\gamma,z)$ and  $U_{\rm K}(\alpha,\gamma,z)$ denote two fundamental solutions\footnote{Standard notations for Kummer's functions are   $M(\alpha,\gamma,z)$ and  $U(\alpha,\gamma,z).$
  We  add the subscript K to avoid confusion with the scaled   velocity $U$.} 
of (\ref{conf.hyp.geom2}).
\end{proposition}
We note that the real parameter $A$ does not have to be positive in the general solutions above.
This will be important when we consider a backward self-similar solution below.

{\bf Proof}\\
This is done by straightforward calculations.
Taking  $A\textcolor{black}{(=a/\nu)}=1$ without loss of generality, consider
$\Psi=\xi e^{-\frac{\xi^2}{2}} w \left(1-\dfrac{B}{2},\dfrac{3}{2},z\right),$ with $z=\xi^2/2.$
Direct calculations show
$$\partial_\xi \Psi= (1-\xi^2) e^{-\frac{\xi^2}{2}} w+\xi^2  e^{-\frac{\xi^2}{2}} \partial_z w,$$
and
$$\partial_{\xi\xi} \Psi= (\xi^3-3\xi) e^{-\frac{\xi^2}{2}} w +(3\xi-2\xi^3)  e^{-\frac{\xi^2}{2}}\partial_z w
+\xi^3  e^{-\frac{\xi^2}{2}}\partial_{zz} w.$$
Thus we get
\begin{eqnarray}
\partial_{\xi\xi} \Psi+\xi \partial_\xi \Psi+B\Psi
&=&e^{-\frac{\xi^2}{2}}\left\{\xi^3 \partial_{zz} w+(3\xi-\xi^3)\partial_z w+(B-2)\xi w\right\}\nonumber\\
&=&e^{-\frac{\xi^2}{2}}\xi\left\{\xi^2 \partial_{zz} w+(3-\xi^2)\partial_z w+(B-2) w\right\}.\nonumber
\end{eqnarray}
As $\xi^2=2z,$ we deduce
$$z \frac{d^2 w}{dz^2}+\left(\frac{3}{2}-z\right)\frac{dw}{dz}+\left(\frac{B}{2}-1\right) w.\;\;\square$$
We need to check which option, $M_{\rm K}$ or $U_{\rm K},$ is acceptable for our purpose.
First we check $M_{\rm K}$.
By the asymptotic formulas  for $|z|\to \infty$ $(\Re z >0)$ (see {\bf Appendix B}),
we have as $|\xi| \to \infty$
$$M_{\rm K}\left(1-\frac{B}{2A},\frac{3}{2},\frac{A\xi^2}{2}\right)
\approx \frac{\Gamma(\frac{3}{2})}{\Gamma(1-\frac{B}{2})}e^{A\xi^2/2}
\left(\frac{A\xi^2}{2}\right)^{-\frac{1}{2}(1+\frac{B}{A})} \propto e^{A\xi^2/2}  \xi^{-(1+\frac{B}{A})}.$$
Hence with this choice we obtain
$$\Psi \approx  \xi^{-\frac{B}{A}} \;\;\mbox{and}\;\;  U=-2\nu \partial_\xi \log \Psi \propto \xi^{-1},$$
which is consistent with the boundary behaviour, i.e. the condition $C \ne 0$ above.

On the other hand, for the other solution $U_{\rm K}$ we have as $|\xi| \to \infty$
$$U_{\rm K}\left(1-\frac{B}{2A},\frac{3}{2},\frac{A\xi^2}{2}\right) \approx
\left(\frac{A\xi^2}{2}\right)^{\frac{B}{2A}-1} \propto \xi^{\frac{B}{A}-2},$$
thus in this case we find 
$$\Psi \propto e^{-A\xi^2/2} \xi^{\frac{B}{A}-1}\;\mbox{and}\; U \propto -A\xi +\left(\frac{B}{A}-1\right)\xi^{-1}.$$
We should discard this option because, when $A\ne0$, the corresponding $U$  does not even belong to
$B^0_{1,\infty}(\mathbb{R}^1),$  due to the presence of the linear  term in $\xi$. 

\begin{figure}[ht]
\begin{minipage}{0.55\linewidth}
  \includegraphics[scale=0.4,angle=0]{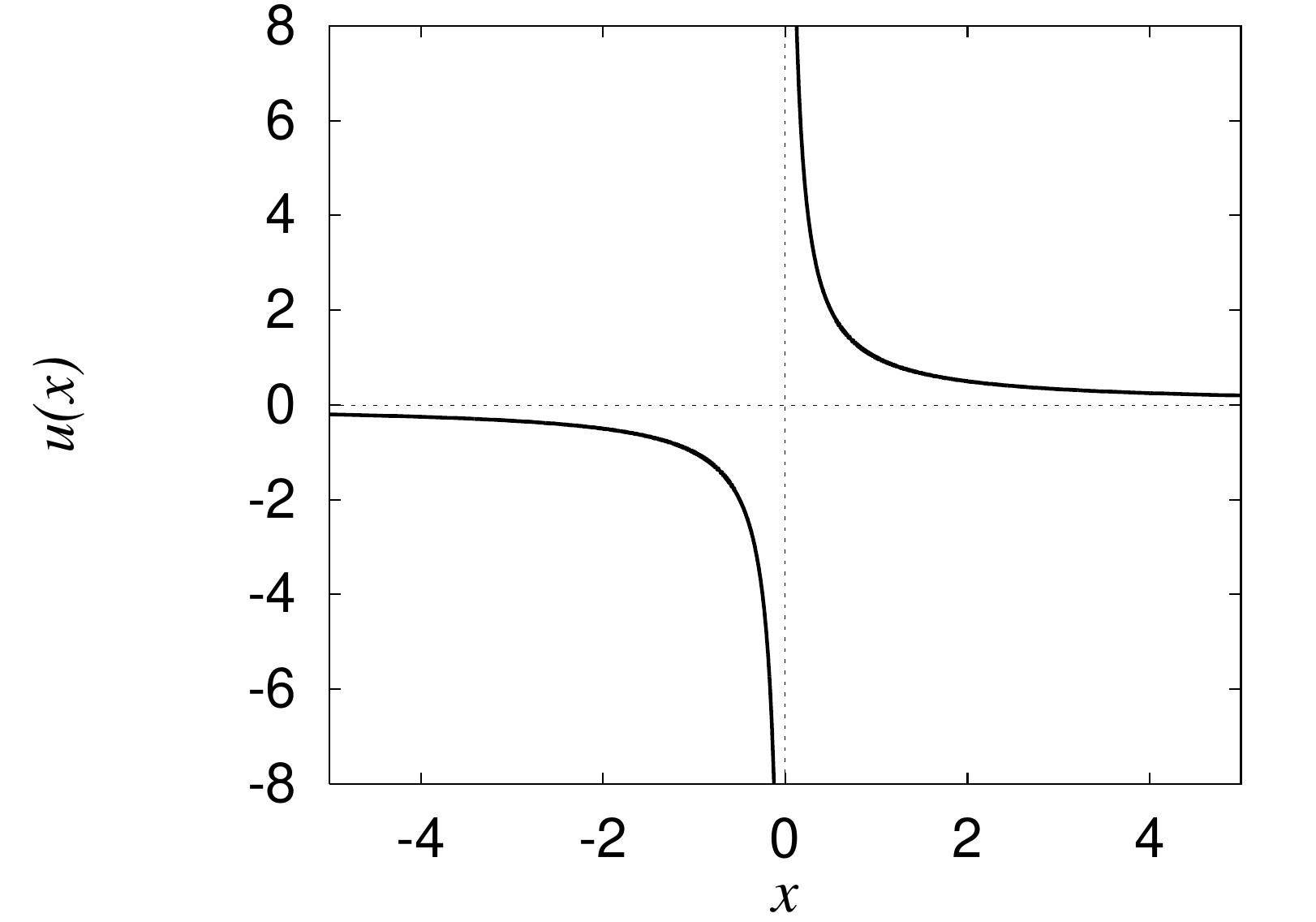}
\caption{The self-similar initial data \textcolor{black}{$u(x)=\frac{1}{x}$ at $t = 0$.}}
\label{IC_kink}
\end{minipage}
\begin{minipage}{0.5\linewidth}
  \includegraphics[scale=0.25,angle=0]{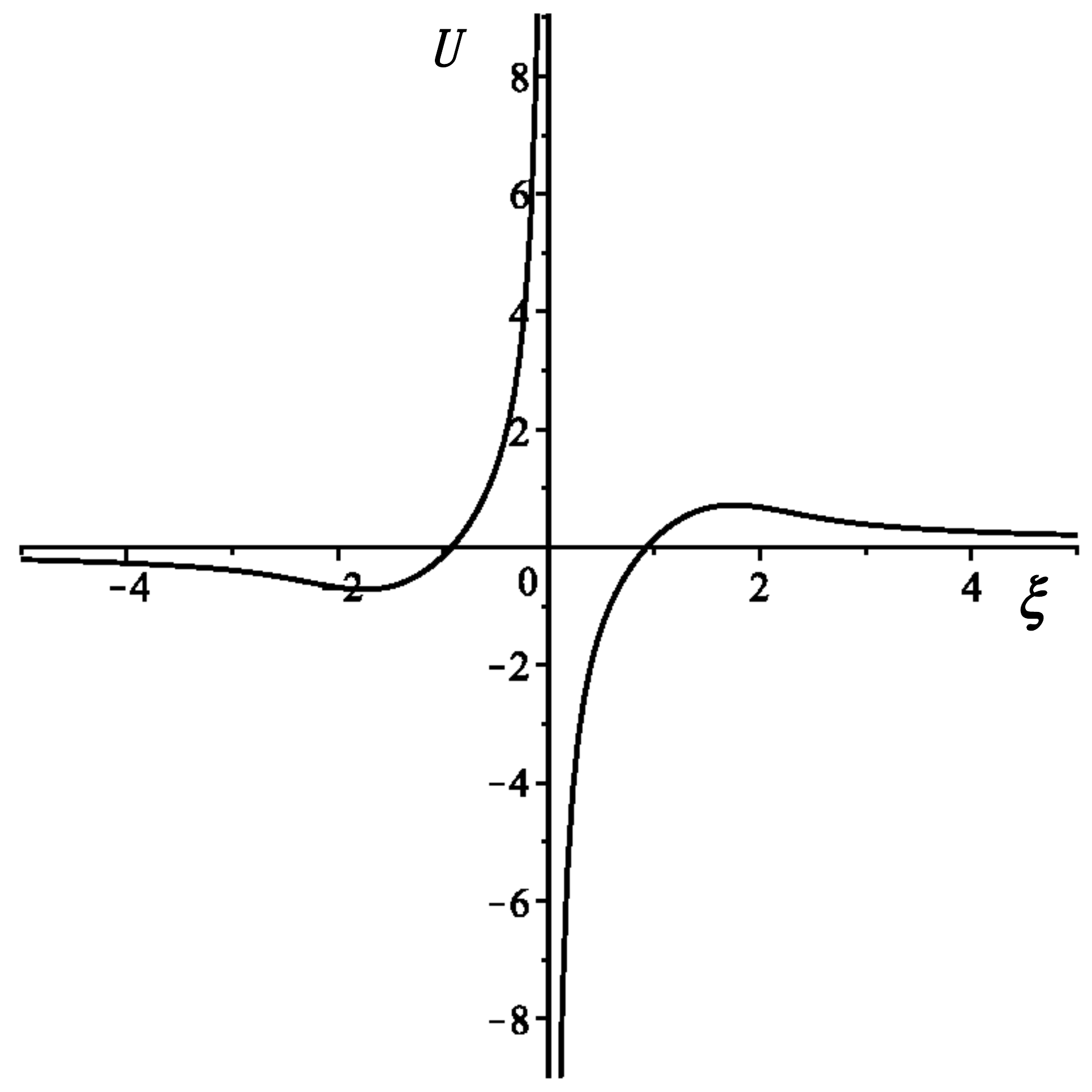}
  \caption{The kink-type solution as $\tau \to \infty$;
    $U(\xi)=-\frac{D'(\xi)}{D(\xi)}=-\frac{1-2\xi D(\xi)}{D(\xi)}.$}
  \label{Burg_kink}
\end{minipage}
\end{figure}

We conclude that for the kink solution, we should choose
$$\Psi=C_1 \xi\exp\left(-\frac{a}{2\nu}\xi^2\right) M_{\rm K}\left(1-\frac{C}{2a},\frac{3}{2},\frac{a}{2\nu}\xi^2\right).$$
It may be in order to have a look at a specific example of the class of solutions.
Replacing $z \to iz,$ in the \textcolor{black}{following} identity, e.g. \cite{OLBC2010},
$$M_{\rm K}\left(\frac{1}{2}\frac{3}{2},-z^2\right)
=\frac{\sqrt{\pi}}{2z} {\rm erf}(z),$$
we have
$$M_{\rm K}\left(\frac{1}{2}\frac{3}{2},z^2\right)
=\frac{\sqrt{\pi}}{2iz} {\rm erf}(iz)
=\frac{\sqrt{\pi}}{2z} {\rm erfi}(z).$$
Here ${\rm erf}(z)=\frac{2}{\sqrt{\pi}}\int_0^z e^{-t^2}dt$
denotes the error function and its imaginary version
${\rm erfi}(z)=\frac{2}{\sqrt{\pi}}\int_0^z e^{t^2}dt.$
\textcolor{black}{In order to make use of the identity, we take $C/a=1$.}
With this a typical example can be given
$$\Psi=\textcolor{black}{C_1}
\xi \exp\left(-\frac{a}{2\nu}\xi^2\right) M_{\rm K}\left(\frac{1}{2}\frac{3}{2},\frac{a\xi^2}{2\nu}\right)
=\textcolor{black}{C_1}
\sqrt{\frac{2\nu}{a}}D\left(\sqrt{\frac{a}{2\nu}}\xi\right),$$
where $D(x) \equiv e^{-x^2}\int_0^x e^{t^2}dt=\frac{\sqrt{\textcolor{black}{\pi}}}{2}H[e^{-x^2}]$
denotes Dawson's integral and $H[\cdot]$ the Hilbert transform.
See Figs. \ref{IC_kink}, \ref{Burg_kink}, \textcolor{black}{in which  we also assume  $a=1$ and $\nu=1/2$  for simplicity.
Note that the profile in Fig.4 is a steady solution in scaled space, 
which develops from the initial profile in the original variables in Fig.3 under time evolution.}

It can be seen that in higher spatial dimensions (\ref{conf.hyp.geom}) is generalised to
\bel{FP_multi}
\triangle \Psi +\frac{a}{\nu}(\bm{\xi}\cdot\nabla)\Psi+\frac{C}{\nu}\Psi=0.
\ee
However, representation formulas for solutions to (\ref{FP_multi}) are not known.

See {\bf Appendix C} for a motivation or rationale for studying two different kinds of steady solutions.

\subsection{Backward self-similar solutions}
Generally speaking, we talk about forward self-similar solutions to study the decaying process
in the late stage of evolution, whereas talk about  backward self-similar solutions to study
whether  solutions blow up in finite time. Needless to mention, no solutions to the Burgers equation blow up if they start
from well-localised smooth initial data under natural boundary conditions.
Nonetheless, because the backward problem differs from the forward
problem only in the sign of the parameter $a,$ if one of them is obtained it is readily transferable to the other one
by flipping the sign of $a,$ provided that we do not bother their boundary behaviour.

In the spirit of \cite{CP1996} we have studied forward self-similar solutions to the Burgers equation
under the setting of $B^{0}_{1,\infty}(\mathbb{R}^1).$ 
What would happen if we consider the backward self-similar solutions in the enlarged function class, or even broader one ?

\begin{figure}[ht]
\begin{minipage}{0.5\linewidth}
  \includegraphics[scale=0.25,angle=0]{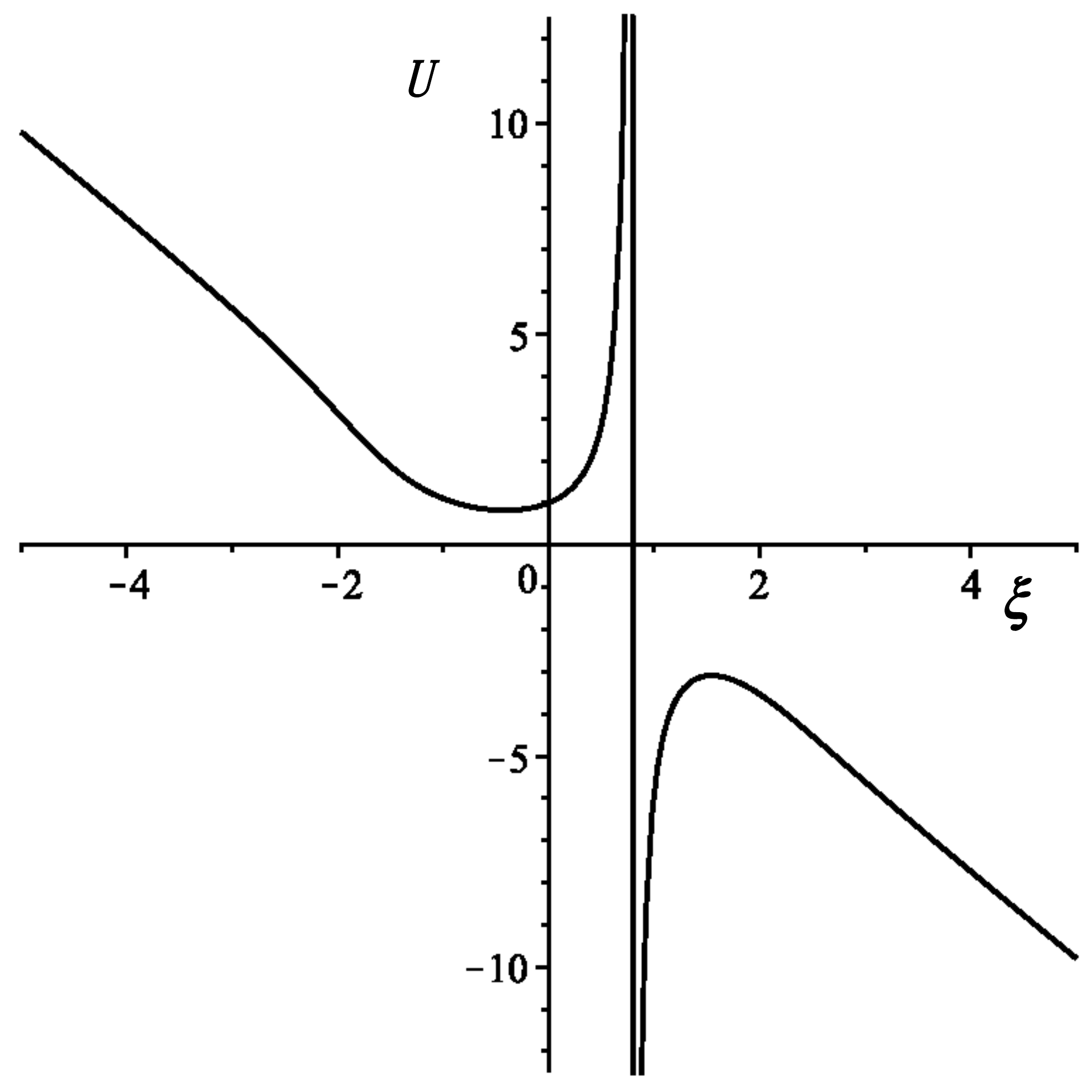}
  \caption{A blowup profile
    $U=\dfrac{e^{\xi^2}}{1-\int_0^\xi e^{\eta^2}d\eta},$
    which behaves $\propto \frac{1}{\xi_*-\xi}$ near some $\xi_*$ and  $\propto \xi$
  as $|\xi| \to \infty.$}
\label{Burg_blowup}
\end{minipage}\begin{minipage}{0.5\linewidth}
\end{minipage}
\end{figure}

We can study 'possible' blow up with backward self-similarity, putting
$$u(x.t)=\frac{1}{\lambda(t)} U\left( \frac{x}{\lambda(t)}\right)$$
with the length-scale $\lambda(t)=\sqrt{2a(t_*-t)},$ \textcolor{black}{where $t_*$ denotes the time of blowup.}
The steady equation reads
$$
U \frac{\partial U}{\partial \xi}
+a \left( \xi  \frac{\partial U}{\partial \xi}+ U \right)
=\nu \frac{\partial^2 U}{\partial \xi^2},
$$
to which the only smooth solution is a trivial one $U \equiv 0$ under natural boundary conditions and smoothness conditions.
Relaxing those conditions, a non-trivial solution  is nonetheless obtained as
$$U(\xi)=\frac{C'e^{a\xi^2/(2\nu)}}{1 -\frac{C'}{2\nu}\int_0^\xi e^{a\eta^2/(2\nu)} d\eta}.$$
This solution is badly-behaved at far distances;\footnote{In Appendix D of \cite{Ohkitani2020} it was stated
  erroneously that $U(\xi) \to \frac{1}{|\xi|}\;\; \mbox{as} \;\; \xi \to \pm\infty,$
  which should be corrected as above.}
 $U(\xi) \propto \xi$ as $\xi \to \pm\infty.$
Furthermore this has a singular point (a pole) somewhere, say at $\xi=\xi_*,$ where
the denominator vanishes.
Thus we have $U(\xi) \propto 1/(\xi-\xi_*)$ around it and is non-integrable $U \notin L^1_{\rm loc}.$
See Fig.\ref{Burg_blowup}.

It is of interest to have another look at the backward self-similar solution.
We first recast the source-type solution
using Kummer's function.
Taking $\beta=1/2$ in the \textcolor{black}{following} identity \cite{OLBC2010} 
$$e^{-z}M_{\rm K}(1, \beta+1,z)=\beta z^{-\beta} \gamma(\beta,z),$$
where $\gamma(\beta,z)\equiv\int_0^z t^{\beta-1} e^{-t} dt,$  
we have
$$e^{-z}M_{\rm K}\left(1, \frac{3}{2},z\right)=z^{-1/2}\int_0^{\sqrt{z}} e^{-s^2}ds.$$
Putting $z=\frac{a}{2\nu}\xi^2,$ we find
$$\Psi=\xi \exp\left(-\frac{a}{2\nu} \xi^2 \right)M_{\rm K}\left(1, \frac{3}{2}, \frac{a}{2\nu} \xi^2 \right)
=\sqrt{\frac{\pi \nu}{2a}} {\rm erf} \left( \sqrt{\frac{a}{2\nu}}\xi\right).$$
When we replace $a \to -a,$ the right-hand side is changed as
$$\Psi \to \frac{1}{i}\sqrt{\frac{\pi \nu}{2a}} {\rm erf} \left( i \textcolor{black}{\sqrt{\frac{a}{2\nu}}}\xi\right)
=\sqrt{\frac{\pi \nu}{2a}} {\rm erfi} \left(\sqrt{\frac{a}{2\nu}}\xi\right),$$
which agrees with backward self-similar solution \textcolor{black}{obtained} above. Note that the asymptotic formula for
$M_{\rm K} (\alpha,\gamma,z)$ for $|z| \to \infty$ does not hold valid for $\alpha=1,$ because $\Gamma(1)=0.$

It may be interesting to consider the following question.
The existence of the forward self-similar (source-type) solutions $\bm{U}$ to the 3D Navier-Stokes
equations is known. Suppose we make a replacement $a \to -a$ in such solutions,
the backward profile $\bm{U}$ must be singular and/or ill-behaved at far distances,
in view of the non-existence of self-similar blowup \cite{NRS1996, Tsai1998}.
We would still be interested in what kind of spatial structure the profile possesses because such a solution
may be helpful in \textcolor{black}{putting constraints under the replacement $a \leftrightarrow -a$  on the sought-after
  forward source-type solutions.}

\section{Self-similar solutions of the Navier-Stokes equation}
\subsection{2D Navier-Stokes equations}
The 2D Navier-Stokes equations  is described by the vorticity equation
 $$\dfrac{\partial \omega}{\partial t}+ \bm{u}\cdot \nabla  \omega                                        
=\nu \triangle \omega.$$
The dynamically-scaled form of the vorticity equation reads
 \bel{NS2D_scaled}
 \dfrac{\partial \Omega}{\partial \tau}+ \bm{U}\cdot \nabla  \Omega                                        
 =\nu \triangle \Omega + a \nabla \cdot (\bm{\xi}\Omega),
 \ee
 \textcolor{black}{
   where  $\Omega(\bm{\xi},\tau)
   =\frac{1}{2at}\omega\left(\bm{\xi},\tau \right)$ denotes the scaled vorticity,
 $\bm{\xi}=\frac{\bm{x}}{\sqrt{2at}}$ and $\tau=\frac{1}{2a}\log t$.} 
Under the assumption of radial symmetry, i.e. the similarity profile $\Omega$ being a function of
$\xi=|\bm{\xi}|$ only, it has a solution
\bel{Burgers_vortex}
\Omega(\xi)=\frac{a \Gamma}{2\pi\nu}\exp\left(-\frac{a|\bm{\xi}|^2}{2\nu} \right),
\ee
where $\Gamma=\int_{\mathbb{R}^2}\Omega(\bm{\xi})d\bm{\xi}$ denotes the circulation.
This is the celebrated Burgers vortex, which is well-localised in space and $\in L^1(\mathbb{R}^2)$.

All the radially-symmetric solutions to the 2D Fokker-Planck equations were worked out explicitly  in \cite{OV2021}.
One of them is the  Burgers vortex above and the other one is lesser-localised in space.
The latter solution reads
\bel{kink_2D}
\Omega(\xi)= \frac{a}{4\nu} \exp\left(-\frac{a \xi^2}{2\nu} \right)
\left({\rm Ei}\left(\frac{a\xi^2}{2\nu}\right) -\log \left(\frac{a}{2\nu} \right) -\gamma  \right),
\ee
where 
${\rm Ei}(x)\equiv -{\rm p.v.}\int_{-x}^{\infty} \frac{e^{-t}}{t} dt$ denotes the exponential integral and
$\gamma \approx 0.577$ the Euler's constant.
\color{black}
Taking $a/2\nu=1$ for simplicity we consider
$$\Omega(\xi)=\frac{1}{2} \exp(-\xi^2 )\left( {\rm Ei}(\xi^2) -\gamma  \right).$$
Recalling
${\rm Ei}(x) =O\left(\dfrac{e^x}{x}\right)\;\;\,\mbox{as}\;\; x \to \infty$
and
${\rm Ei}(x) \approx \gamma +\log x +x \;\;\,\mbox{as}\;\; x \to 0+,$ we note 
its  asymptotic properties as follows.

\begin{itemize}
\item It decays slowly  $\Omega(\xi) \approx \dfrac{1}{2\xi^2}\;\;\mbox{as}\;\; \xi \to \infty.$ Hence
$\Omega(\xi)$ does not belong  to $L^1(\mathbb{R}^2)$, but does belong to the Besov space $B^0_{1,\infty}(\mathbb{R}^2)$.
\item Near the origin it shows a mild singularity as
$\textcolor{black}{\Omega(\xi) \approx  \log \xi}\;\;\mbox{as}\;\; \xi \to 0+,$ that is, it is discontinuous in vorticity at $\xi=0$.

\item It can be verified nonetheless that the corresponding velocity is continuous there.
The azimuthal component velocity is given by
$$\textcolor{black}{U_\theta(\xi) =\frac{1}{\xi}\int_0^{\xi} \eta\Omega(\eta) d\eta
  =\frac{1}{4\xi}\left\{ \log \xi^2-e^{-\xi^2} \left({\rm Ei}(\xi^2)-\gamma \right) \right\}}.$$
It has the following asymptotic behaviours:
$U_\theta(\xi) \approx \frac{\xi}{2}\log \xi \;\;\mbox{as}\;\; \xi \to 0+ \;\mbox{and}\;
U_\theta(\xi) \approx \frac{1}{2\xi}\log \xi \;\;\mbox{as}\;\; \xi \to \infty.$
\end{itemize}
 Reverting to the original variables from (\ref{kink_2D}) and discarding terms associated with the Gaussian solution,
 we obtain a particular solution\footnote{
This is same the procedure by which we get the so-called Lamb-Oseen decaying
 vortex from (3.2).}
$$
\omega(r,t)=\frac{1}{2at}\Omega\left(\frac{r}{\sqrt{2at}}\right)
=\frac{1}{8\nu t}\exp\left(-\frac{r^2}{4\nu t} \right) {\rm Ei}\left(\frac{r^2}{4\nu t} \right).
$$
We have, for fixed $t$,
$$\omega(\bm{x},t) \approx \frac{\log r}{4\nu t}\;\;\mbox{as}\;\; r \to 0+$$
by ${\rm Ei}(x)\approx \log x\;\;\mbox{as}\;\; x \to 0+.$
Hence the property of the vorticity being singular  at the origin persists throughout
time evolution. 
We also have
$$\omega(x,t) \approx \frac{1}{2r^2}\;\;\mbox{as}\;\; t \to 0+.$$
The corresponding initial data is exactly scale-invariant as expected.

The azimuthal velocity field in the original variables is given by
$$u_\theta(r,t)=\frac{1}{r}\int_0^r s \omega(s,t)ds$$
$$=\frac{1}{4r}\left\{ \log\left(\frac{r^2}{4\nu t}\right)-\exp\left(-\frac{r^2}{4\nu t}\right)
{\rm Ei} \left(\frac{r^2}{4\nu t}\right) +\gamma \right\}.$$
Its asymptotic properties are
$$u_\theta(r,t) \approx \frac{\log r}{2r}\;\mbox{as}\;t \to 0+\;\mbox{and}\;
u_\theta(r,t) \approx \frac{r}{8\nu t}\log r\;\mbox{as}\; r \to 0+.
$$
\color{black}

The second \textcolor{black}{non-Gaussian} solution may serve as a \textcolor{black}{replacement for the} 'kink-type' solution to the problem. However,
neither  realisability nor stability  of (\ref{kink_2D}) is known.

\subsection{3D Navier-Stokes equations}
As already noted, in three dimensions the existence of the forward self-similar solutions is known, e.g.
\cite{CP1996, JS2014},
but the precise functional form of the  solution is not.
If the initial condition is well-localised,
we know that it takes a near-Gaussian form
\textcolor{black}{if the dependent variable is chosen suitably,}
i.e. as the vorticity curl in three-dimensions.

\color{black}
Using the vorticity curl $\bm{\chi}=\nabla \times \bm{\omega}$ the governing equations read
$$\frac{\partial \bm{\chi}}{\partial t}
=\triangle( \bm{u} \cdot \nabla \bm{u}+ \nabla p)  +\nu \triangle  \bm{\chi},$$
where $\bm{u}$ denotes the velocity and 
Under the dynamic scaling transformations
$$\bm{\xi}=\frac{\bm{x}}{\sqrt{2 at}},\;\;                                                                       
\tau=\frac{1}{2 a}\log t,
\bm{\chi}(\bm{x},t)=\frac{1}{(2at)^{3/2}}\bm{X}(\bm{\xi},\tau),$$
we find
$$\frac{\partial \bm{X}}{\partial \tau}
=\triangle \left(\bm{U}\cdot\nabla\bm{U}+\nabla P \right)
+\nu \triangle  \bm{X}+a\nabla\cdot(\bm{\xi}\otimes \bm{X}).
$$
\color{black}
A perturbative attempt  of the determination of source-type solutions can be found in \cite{OV2021}.
The leading-order approximation solution is explicitly given  by a Gaussian function {\textit{modulo} incompressibility
and the corrections due to the nonlinear term is estimated to be small.   The conjugate solution at leading-order is
also worked out explicitly.

\section{Lifting of self-similar solution to more general ones}
We discuss what  can be learnt from studying self-similar solutions and identify  open problems in this regard.
 
\subsection{1D Burgers equation}
When we recast (\ref{Burgers_source1}) as
\bel{Burgers_source2}
U(\xi)=-2\nu \frac{\partial}{\partial \xi}\log\left(
{1 -\displaystyle{\frac{C'}{2\nu}\int_0^{\xi}}   \exp \left( -\frac{a \eta^2}{2 \nu}\right) d\eta}
\right),
\ee
it is reminiscent of the structure of the celebrated Cole-Hopf transform. In other words, the source-type solution
encodes the vital information as to how we may linearise this nonlinear equation.

\color{black}
We don't intend to add anything new to the understanding of the Burgers equation,
rather we will show how we can recover the Cole-Hopf transformation on the
basis of a particular self-similar solution (viz. the source type solution).

Let us pretend that we don't know the Cole-Hopf transformation. It's still straightforward
to derive the source-type solution, which is  slightly non-Gaussian.  It is then
possible to recover the Cole-Hopf linearisation by replacing the self-similar
heat flow with a general heat flow, which we call the lifting procedure.

For illustration we will have a look at the details of the procedure.
Assume that the velocity potential $\Phi(\xi)$ is given by
$$\Phi(\xi)=f(\widehat{\Psi}),$$
where $f$ is a function of
the self-similar source type solution of the heat equation
$\widehat{\Psi}=\int_0^{\xi}\exp \left( -\frac{a \eta^2}{2 \nu}\right) d\eta$.
Then we have
$$U=\partial_{\xi}\Phi=f'(\widehat{\Psi})\partial_{\xi}\widehat{\Psi}.$$
Writing
$$U(\xi)=F(\widehat{\Psi}; \partial_{\xi}\widehat{\Psi}),$$
we have trivially
$$F(x;y)=y f'(x),$$
and the function $F(x;y)$ satisfies the following scaling property
\bel{scaling1D}
F(x;\alpha y)=\alpha F(x;y), \;\mbox{for} \;\forall\; \alpha >0.
\ee
It should be noted that hereafter the arguments $x,y,$ etc. in the function $F$ do not represent
spatial coordinates.
\color{black}

The procedure of spotting (or, recovering, in this case)
more general solutions can be formalised in the following steps.

{\bf Step 1.} Assume a self-similar profile is obtained in the form
\bel{step1_1D}
U(\xi)=F( \widehat{\Psi}(\xi);\partial_\xi \widehat{\Psi}(\xi)) \equiv \frac{\partial_\xi \widehat{\Psi}}{1-\frac{1}{2\nu}\widehat{\Psi}}.
\ee
The profile $U(\xi)$ is a \textit{near-identity} transformation of the last argument of the function, i.e.
that of the Gaussian function $\partial_\xi \widehat{\Psi}(\xi)$.
Luckily for the Burgers equation $F(\cdot;\cdot)$ is known explicitly $F(x;y)=\frac{y}{1-x/(2\nu)},$
as indicated by the symbol $\equiv$  above.
By definition, a particular solution is obtained by reverting to the original variables:
$$u(x,t)=\frac{1}{\sqrt{2at}}
F\left(\widehat{\Psi}\left(\frac{x}{\sqrt{2at}}\right);\partial_\xi \widehat{\Psi}\left(\frac{x}{\sqrt{2at}}\right)\right)
\equiv \frac{1}{\sqrt{2at}}\frac{\partial_\xi \widehat{\Psi}(\frac{x}{\sqrt{2at}})}{1-\frac{1}{2\nu}\widehat{\Psi}(\frac{x}{\sqrt{2at}})}.$$

{\bf Step 2.} Replacing the self-similar heat solution with the general heat flow, we obtain
a more general class of solutions
\begin{eqnarray}
u(x,t)&=&\frac{1}{\sqrt{2at}}
 F\left(\widehat{\psi}\left(x,t\right); \sqrt{2at}\partial_x \widehat{\psi}\left(x,t\right)\right) \nonumber\\
&=&F\left(\widehat{\psi}\left(x,t\right);\partial_x \widehat{\psi}\left(x,t\right)\right)
 \equiv  \frac{\partial_x \widehat{\psi}(x,t)}{1-\frac{1}{2\nu} \widehat{\psi}(x,t)},\nonumber
\end{eqnarray}
where
$$
\widehat{\psi}(x,t)=\frac{1}{\sqrt{4\pi\nu t}}\int_{-\infty}^{\infty} \widehat{\psi}(y,0) \exp\left(-\frac{(x-y)^2}{4\nu t} \right)dy
$$
denotes the \textcolor{black}{general} heat flow.
Note that the \textcolor{black}{length scales} $\sqrt{2at}$ cancel out because of (\ref{scaling1D}).
In the case of the Burgers equation, this last form  provides the general solution.

Before closing this subsection it may be in order to emphasise the  following fact.
In terms of the scaled velocity potential $\Phi(\xi)$ the exact solution is given by
$\Phi=-2\nu \log \left(1-\frac{1}{2\nu}\widehat{\Psi}\right),$
whose leading-order approximation agrees with the scaled heat flow $\Phi(\xi) \approx \widehat{\Psi}(\xi).$
This confirms the near-identity nature of $\Phi(\xi)$.

\subsection{2D Navier-Stokes equations}
\textcolor{black}{In 2D incompressible flows}
the source-type solution is given by the Burgers vortex, which is not only near-Gaussian, but also \textit{exactly}
 Gaussian. Because of this peculiarity,
 the final steady state does not contain any useful information regarding the nonlinear terms.  
 Hence it is impossible to lift (or generalise) the final state to find a more general class of solutions
\textcolor{black}{and we end up with obtaining the linearised solution only.}
 Curiously enough  the better understood 2D Navier-Stokes equations defy the current approach  to gain
 some information about  their solutions.

 \textcolor{black}{Regarding this, it is in order to include a bit more detailed description based on kinematic relationship.
 The stream function corresponding to (\ref{Burgers_vortex}), with $\Gamma=1$, is given by $\widehat{\Psi}(\xi)
 =\dfrac{1}{4\pi} {\rm Ei}\left(-\frac{a|\bm{\xi}|^2}{2\nu} \right),$
 which agrees with the scaled heat flow in two dimensions.
 Assume that the scaled stream function $\Psi(\bm{\xi})$ is given by $\Psi(\bm{\xi})=f(\widehat{\Psi}(\xi)).$
 It is readily derived
 $$\Omega(\bm{\xi})=-\triangle \Psi=  -\triangle \widehat{\Psi} f'(\cdot) -|\nabla \widehat{\Psi}(\xi)|^2 f''(\cdot).$$
 Writing $\Omega(\bm{\xi})=F\left(\widehat{\Psi}(\xi), \nabla \times\widehat{\Psi}(\xi),-\triangle \widehat{\Psi}(\xi)\right),$
 we find $$F(x,y;z)=f'(\cdot)z-f''(\cdot)y^2.$$ The function $F$ satisfies the following scaling
 $$F(x,\alpha y; \alpha^2 z)=\alpha^2 F(x,y;z),\;\mbox{for} \; \forall \; \alpha >0.$$
 The only solution we know is the identity, that is, $f(x)=x.$
}
 \subsection{3D Navier-Stokes equations}
 We take the vorticity curl, $\bm{\chi}=\nabla \times \bm{\omega}$ as the basic dependent variable,
 whose dynamically-scaled version is denoted  by $\bm{X}(\bm{\xi}).$
 \textcolor{black}{With this choice, the linearised equations have the Fokker-Planck operator}
In view of \textcolor{black}{the critical scale-invariance of type 2}  \cite{OV2021} it is most convenient for our analysis,
as the leading-order approximation
is basically given by the Gaussian function, i.e. the Gaussian function \textit{modulo} incompressibility.

The steady version of the dynamically-scaled Navier-Stokes equations reads \cite{OV2021}
\bel{NS3D_scaled}
\triangle^{*} \bm{X} \equiv
\triangle \bm{X} +\frac{a}{\nu} \nabla\cdot(\bm{\xi}\otimes \bm{X})
=-\frac{1}{\nu}\triangle\mathbb{P} \left(\triangle^{-1}\bm{X}\cdot\nabla \triangle^{-1}\bm{X}\right).
\ee
By a formal analysis we show the following
\begin{proposition}
  The successive approximations to \textcolor{black}{the solution $\bm{X}$ of} the equations (\ref{NS3D_scaled}) are given by a functional
 \footnote{It is a functional rather than a function as we need to take into account nonlocal interactions due
   to the incompressible condition.}  of
 \textcolor{black}{
 $$\{(\nabla \times)^k \widehat{\bm{\Psi}}(\bm{\xi})\;|\;  k=0,1,2,3\}.$$
 Here $(\nabla \times)^3 \widehat{\bm{\Psi}}(\bm{\xi})=\mathbb{P}\bm{M}G,$
  defined with $\bm{M}=\int \bm{X}(\bm{\xi})d \bm{\xi},$
  $G=\left(\frac{a}{2\pi \nu} \right)^{3/2}\exp(-\frac{a}{2\nu}|\bm{\xi}|^2)$
  and $\mathbb{P}$  solenoidal projection such that $\nabla \cdot \widehat{\bm{\Psi}}(\bm{\xi})=0$.}
\end{proposition}
Note that $\widehat{\bm{\Psi}}(\bm{\xi})$ denotes the scaled heat flow, whereas $\bm{\Psi}(\bm{\xi})$
the scaled vector potential.

{\bf Proof}\\
We consider the following expansion for small $\epsilon >0,$
$$\bm{X}(\bm{\xi})=\epsilon \bm{X}_1+\epsilon^2 \bm{X}_2+\epsilon^3 \bm{X}_3+\ldots,$$
deferring the justification of smallness of  $\epsilon$ by its identification as the Reynolds number.
We  prove by mathematical induction that 
each $\bm{X}_n$ can be represented by \textcolor{black}{a combination of  functionals of $\widehat{\bm{\Psi}}(\bm{\xi}),$
including e.g. $\triangle^{-1}\widehat{\bm{\Psi}}$  and the derivatives of $\widehat{\bm{\Psi}}(\bm{\xi})$.}
Equating the terms  with the same powers in $\epsilon,$ we derive equations for $\bm{X}_n$ for $n \geq 1.$

(i) We will first confirm that this is the case for $n=1$.\\
To leading order at $O(\epsilon)$ we have
$$
\triangle^* \bm{X}_1=0,$$
from which it follows that $\bm{X}_1=\mathbb{P}\bm{M}G.$
Indeed $\bm{X}_1$ is a functional of the desired form.
By definition $\bm{X}_1=(\nabla \times)^3 \bm{\Psi}_1(\bm{\xi}),$
we also note that the leading-order approximation satisfies
$\bm{\Psi}_1(\bm{\xi})=\widehat{\bm{\Psi}}(\bm{\xi}).$

(ii)  Assuming that the statement holds up to  step $k (\leq n),$
we will deduce that the it also holds for step $(n+1).$ \textcolor{black}{For illustration let us first take a look at,
 for example,   $O(\epsilon^2)$ and $O(\epsilon^3)$.} To next-to-leading order at $O(\epsilon^2)$ we have
$$\triangle^* \bm{X}_2
=-\frac{1}{\nu}\triangle\mathbb{P}(\triangle^{-1}\bm{X}_1 \cdot \nabla \triangle^{-1}\bm{X}_1)$$
and at the third order $O(\epsilon^3)$
$$
\triangle^* \bm{X}_3
=-\frac{1}{\nu}\triangle\mathbb{P}(\triangle^{-1}\bm{X}_1 \cdot \nabla \triangle^{-1}\bm{X}_2
+\triangle^{-1}\bm{X}_2 \cdot \nabla \triangle^{-1}\bm{X}_1).$$
Likewise, at $O(\epsilon^{n+1})$ for $\forall n \in \mathbb{N},$ we have
$$\triangle^* \bm{X}_{n+1}=-\frac{1}{\nu}\triangle\mathbb{P}\sum_{l=1}^{n}(\triangle^{-1}\bm{X}_l \cdot \nabla \triangle^{-1}\bm{X}_{n+1-l}),$$
or
\bel{next_step}
\bm{X}_{n+1}=-(\nu\triangle^*)^{-1}\triangle\mathbb{P}\sum_{l=1}^{n}(\triangle^{-1}\bm{X}_l \cdot \nabla \triangle^{-1}\bm{X}_{n+1-l}).
\ee
Here use has been made of the inverse Fokker-Planck  operator is given by \cite{OV2021}
$$(\nu \triangle^*)^{-1} \equiv -\int_{0}^{\infty} ds e^{\nu s \triangle^*} =\int d\bm{\eta} g(\bm{\xi},\bm{\eta})$$
with
$$g(\bm{\xi},\bm{\eta}) \equiv \frac{-1}{(2\pi\nu)^{3/2}} {\rm f.p.}\int_{\sqrt{a}}^{\infty}                                  
\frac{\sigma^2 d\sigma}{\sigma^2-a}e^{-\frac{1}{2\nu}|\sigma \bm{\xi} -\bm{\eta}\sqrt{\sigma^2-a}|^2},$$
where f.p. denotes Hadamard's finite part. Noting that $(\triangle^*)^{-1}\triangle \mathbb{P}$
represents an integral operator of
order zero, we  see that the right-hand side of (\ref{next_step}) is a functional of $\bm{X}_k$'s $(1\leq k \leq n).$
Because they are all represented as functionals of $\widehat{\bm{\Psi}}(\bm{\xi})$
and its derivatives, so is $\bm{X}_{n+1}$.
Hence we deduce that $\bm{X}_n$ \textcolor{black}{can be expressed as} a functional of $\widehat{\bm{\Psi}}(\bm{\xi})$ and its derivatives
for all $n \in \mathbb{N}.$ $\;\;\square$

\textcolor{black}{
We note that that the nonlinear contribution $\bm{X}_n\; (n \geq 2) $ has the second order derivative of $\bm{\Psi}_1(\bm{\xi})$
at most and the $(5-2n)$-th order derivative at least. At $n=3$ already we have $5-2n < 0$ and 
this means that it may involve integrals of $\bm{\Psi}_1(\bm{\xi})$, hence $F$
may be a functional, rather than a function.}
We also note that when the $\epsilon$-expansion is uniformly convergent, $\bm{X}$ itself is  a functional of
$\widehat{\bm{\Psi}}(\bm{\xi})$ and its derivatives.
To see how we can make $\epsilon$ arbitrarily small, take for example, the next-to-leading order approximation
\textcolor{black}{(iteration)}
\cite{OV2021}
$$
\bm{X}=\bm{X}_1
-\frac{1}{\nu}\mathbb{P} \left(\triangle^{-1}\bm{X}_1\cdot\nabla \triangle^{-1}\bm{X}_1\right)
$$
\bel{next2leading}
=\mathbb{P}\bm{M}G
-\frac{1}{\nu}\mathbb{P} \left(\triangle^{-1}  \mathbb{P}\bm{M}G\cdot\nabla \triangle^{-1}\mathbb{P}\bm{M}G \right).
\ee
Introducing the following  variables for non-dimensionalisation
$\widetilde{\bm{X}}=\bm{X}/\nu\,\mbox{and}\,\widetilde{\bm{M}}=\bm{M}/\nu,$
we find
$$\widetilde{\bm{X}}=
\mathbb{P}\widetilde{\bm{M}}G
-\mathbb{P}\left(\triangle^{-1}\mathbb{P}\widetilde{\bm{M}}G\cdot\nabla\triangle^{-1}\mathbb{P}\widetilde{\bm{M}}G \right),$$
where  $Re=|\widetilde{\bm{M}}|=|\bm{M}|/\nu$ denotes the Reynolds number.
Identifying $Re=\epsilon,$ this corresponds to the $O(\epsilon^2)$ approximation. We can argue similarly for the
$O(\epsilon^n)$  approximations as well for  $\forall n \in \mathbb{N}$. 

On the basis of Proposition 4.1,
we consider a self-similar profile in the following form\footnote{Care should be taken of the notations;
  it means that  $\bm{F}$ is a functional of $\widehat{\bm\Psi}$ and its derivatives with
  $(\nabla \times)^3 \bm{\widehat{\Psi}}$ at the highest.}
\bel{profile_3D}
\bm{X}(\bm{\xi})=\bm{F} \left(  \bm{\widehat{\Psi}}(\bm{\xi}),
\nabla \times \bm{\widehat{\Psi}}(\bm{\xi}),
(\nabla \times)^2 \bm{\widehat{\Psi}}(\bm{\xi});
  (\nabla \times)^3 \bm{\widehat{\Psi}}(\bm{\xi})\right)
  \ee
  for some functional $\bm{F},$  
  where $\bm{\widehat{\Psi}}$ denotes the scaled heat flow,
  $(\nabla \times)^3\bm{\widehat{\Psi}}=\mathbb{P}\bm{K}\widehat{G},$ 
  $\bm{K}$ is a function of $\bm{M}=\int \bm{X}(\bm{\xi})d \bm{\xi}$
and $\widehat{G}=\exp(-\frac{a}{2\nu}|\bm{\xi}|^2).$
The functional $F(\bm{x},\bm{y},\bm{z};\bm{w})$ is a near-identity transformation of the last argument $\bm{w}$ (the solenoidal Gaussian function) and 
satisfies
\bel{scaling3D}
\bm{F}(\bm{x}, \alpha \bm{y}, \alpha^2 \bm{z}; \alpha^3 \bm{w})=\alpha^3 \bm{F}(\bm{x},\bm{y},\bm{z};\bm{w}) \;\mbox{for} \;\forall \alpha >0.\ee
The profile (\ref{profile_3D}) is a rather strong assumption, even though we take it approximately. If $\bm{F}$ is available, at least
part of solutions to the Navier-Stokes equations is reducible to those of the heat equations.
Unfortunately, the precise form of $\bm{F}$ is not known at the moment, but we know that it is close to the Gaussian
and how close it is \cite{OV2021}. \textcolor{black}{Assuming that $F$ is obtained, the procedure goes as follows.}

{\bf Step 1.} When the self-similar profile $\bm{F}$ is given explicitly,
a particular self-similar solution is obtained as
$$\bm{\chi}(\bm{x},t)=\frac{1}{(2at)^{3/2}}
  \bm{F}\left(
  \bm{\widehat{\Psi}}\left(\frac{\bm{x}}{\sqrt{2at}}\right),
   \nabla \times \bm{\widehat{\Psi}}\left(\frac{\bm{x}}{\sqrt{2at}}\right),
   (\nabla \times)^2 \bm{\widehat{\Psi}}\left(\frac{\bm{x}}{\sqrt{2at}}\right);
   (\nabla \times)^3 \bm{\widehat{\Psi}}\left(\frac{\bm{x}}{\sqrt{2at}}\right)
  \right).$$                                                                                          

{\bf  Step 2.}  We seek a  more general class of solution on this basis
$$\bm{\chi}(\bm{x},t) = \frac{1}{(2at)^{3/2}}
  \bm{F} \left(\bm{\widehat{\psi}}(\bm{x},t), (2at)^{1/2}\,\nabla \times\bm{\widehat{\psi}}(\bm{x},t),
  2at\, (\nabla \times)^2 \bm{\widehat{\psi}}(\bm{x},t);
  (2at)^{3/2}\, (\nabla \times)^3 \bm{\widehat{\psi}}(\bm{x},t) \right).$$
  The first three arguments  give rise to a factor of $(2at)^{3/2}$ in total, as a result of scaling (\ref{scaling3D})
  and hence we have
  $$\bm{\chi}(\bm{x},t)=
  \bm{F}\left(
  \bm{\widehat{\psi}}(\bm{x},t),  \nabla \times \bm{\widehat{\psi}}(\bm{x},t),
   (\nabla \times)^2 \bm{\widehat{\psi}}(\bm{x},t);
  (\nabla \times)^3 \bm{\widehat{\psi}}(\bm{x},t)\right). $$
It is not known how general a class of functions such a construction can cover.

To see how cancellations take place,
it is helpful to consider the following example  based on the next-to-leading order approximation.
  On the right-most side of (\ref{next2leading}), the first term gives rise to $(2at)^{3/2}$
  so does the second $(2at)^{1/2}\cdot (2at)=(2at)^{3/2}.$
  
\textcolor{black}{
  {\bf Remark}\\
  In three dimensions, under the assumption of $\bm{\Psi}=\bm{f}( \bm{\widehat{\Psi}}),$
  a simpler form of $\bm{F}$   can be obtained using spatial derivatives rather than curls.
  Write
  $$\bm{X}=\widetilde{\bm{F}}\left( \bm{\widehat{\Psi}}, D\bm{\widehat{\Psi}}, D^2\bm{\widehat{\Psi}}; D^3\bm{\widehat{\Psi}}
  \right),$$
  where $D_i=\partial_{\xi_i}\;\;i=1,2,3$ stands for any spatial derivatives.
  After some algebra we find explicitly
  $$X_i=-\epsilon_{ikq}\left\{
  \frac{\partial s_j}{\partial \xi_k}\frac{\partial s_m}{\partial \xi_l}\frac{\partial s_n}{\partial \xi_l}
  \frac{\partial^3 f_q}{\partial s_j\partial s_m \partial s_n}
  +\left( \frac{\partial s_j}{\partial \xi_k}\triangle s_m
  + 2  \frac{\partial^2 s_j}{\partial \xi_k\partial \xi_l} \frac{\partial s_m}{\partial \xi_l}  \right)
  \frac{\partial^2 f_q}{\partial s_j\partial s_m}
  +\frac{\partial \triangle s_j}{\partial \xi_k}\frac{\partial f_q}{\partial s_j} 
  \right\}$$
  where we have put $s_i=\widehat{\Psi}_i(\bm{\xi})$ for simplicity.
Observe that the above expression reduces to $\bm{X}=-\nabla \times \triangle \widehat{\bm{\Psi}}.$
for the identity transformation $\frac{\partial f_q}{\partial s_j}=\delta_{qj}.$ }

  \section{Summary and outlook}

 In this paper the following things are done.
 As part of forward self-similar solutions to the Burgers equation
 we have identified a class of kink-type solutions, on top of the well-known source-type solutions.
They are given by Kummer's confluent hypergeometric function $M_{\rm K}$.
 We also noted  ill-behaved 'blow-up' profiles by flipping the sign of the parameter $a.$

 We discussed the Navier-Stokes equations; \textcolor{black}{in two dimensions we discussed a self-similar profile 
which may be regarded as a conjugate to the Burgers vortex.
Some of its asymptotic properties have been studied, whereas clarification of its significance
it may play, e.g. its stability as a dynamical system, requires further investigation.}

 \textcolor{black}{ 
We also discussed applications of the self-similar solution in three dimensions.
Some properties of the self-similar profile have been analysed formally and
possible lifting to more general class of solutions, at least approximately, is suggested.
 For the final topic, it may be worthwhile to try computing approximate solutions by numerical methods.
 This is also left  for future study.}

\appendix
\section{Riccati equation}
Consider the Riccati's ordinary differential equation
$$\frac{dy}{dx}=a(x)y^2+b(x)y+c(x),$$
where $a(x), b(x)$ and $c(x)$ are given functions.
It is known that a substitution of the following form
$$y=-\frac{u'(x)}{a(x)u(x)}$$
reduces the above to a linear second-order homogeneous equation
$$\frac{d^2 u}{dx^2}-\left(\frac{a'(x)}{a(x)}+b(x)\right)\frac{du}{dx}+a(x)c(x)u=0.$$

\section{Confluent hypergeometric equation}
Kummer's confluent hypergeometric equation \cite{OLBC2010}
$$z\frac{d^2 w}{dz^2}+\left(\gamma-z\right)\frac{dw}{dz}-\alpha w=0$$
has two fundamental solutions $w(z)=M_{\rm K}(\alpha,\gamma,z)$ and $U_{\rm K}(\alpha,\gamma,z)$.
Their asymptotic behaviours are given as follows.

For $\Re(z) >0,$ as $|z|\to \infty$ we have
$$\left\{
\begin{array}{l}
  M_{\rm K}(\alpha,\gamma,z)\approx \dfrac{\Gamma(\gamma)}{\Gamma(\alpha)}e^z z^{\alpha-\gamma},
  (\alpha \ne 1, \gamma \ne - n), \\
   \noalign{\vskip0.2cm}
  U_{\rm K}(\alpha,\gamma,z)\approx z^{-\alpha}, 
\end{array}
\right. $$
where $\Gamma(\cdot)$ denotes the gamma function.

Also, as $|z|\to 0$ we have
$$\left\{
\begin{array}{ll}
M_{\rm K}(\alpha,\gamma,z)= 1,\;\; & (\gamma \ne - n), \\
  \noalign{\vskip0.2cm}
U_{\rm K}(\alpha,\gamma,z)= \dfrac{\Gamma(\gamma-1)}{\Gamma(\alpha)}z^{1-\gamma} + O(1),\;\; &(\alpha \ne 1, 1 < \Re(\gamma)<2),
\end{array}
\right. $$
where $n \in \mathbb{N}$.
\section{Source-kink duality}

The PDE, Burgers equation,  allows at least in two different kinds of
approximations based on ODEs via particle systems.
Those particle pictures are well-known, but best stated here for motivation.

One is \\
(1) Propagation of Wiener's chaos, e.g.\cite{Sznitman1991}
$$dX_i=\frac{1}{2N}\sum_{j\ne i}b(X_j-X_i) dt +dW_i,$$
where $dW$ denotes Brownian motion, the drift velocity
$b=\delta$ for the Burgers equation. This is related with the source-type solution.

The other one is\\
(2) Pole decomposition, e.g.\cite{Aref1983}
$$u(x,t)=-2 \nu \sum_{j=1}^N \frac{1}{x-z_j(t)},$$
$$\frac{d z_j}{dt}=-2\nu \sum_{k=1,k\ne j}^{N}\frac{1}{z_j-z_k},$$
where $z_j$ denotes the locations of poles in the complex plane.
This is related with the  kink-type solution, represented by Kummer's $M_{\rm K}$.

A crude explanation why we have two different views is as follows.
There are two fundamental solutions to the Fokker-Planck equation; the Gaussian function and the Dawson's integral, which are related by the Hilbert transform to each other.
One of them converges to the Dirac mass $\delta$ and the other to a Cauchy kernel $1/z,$
respectively in suitable limits.


\enlargethispage{20pt}









\end{document}